\documentclass{article}

\usepackage{arxiv}
\makeatletter
\AtBeginDocument{%
  \pagestyle{fancy}%
  \rhead{}%
}

\renewcommand{\keywords}[1]{\par\noindent\textbf{Keywords:} #1}
\makeatother
\usepackage{booktabs}
\usepackage{tabularx}
\usepackage{tikz}
\usepackage{tikz-3dplot}
\usepackage[utf8]{inputenc} 
\usepackage[T1]{fontenc}    
\usepackage{hyperref}       
\usepackage{url}            
\usepackage{amsfonts}       
\usepackage{nicefrac}       
\usepackage{microtype}      
\usepackage{lipsum}
\usepackage{graphicx}
\usepackage{enumitem}
\usepackage{subcaption}
\usepackage{amsmath}
\usepackage{array} 
\usepackage{natbib}
\usepackage{lineno}         

\usepackage{setspace}
\usetikzlibrary{arrows.meta,positioning,calc}
\usepackage{pgfplots}
\pgfplotsset{compat=1.18}

\usepackage{authblk}

\makeatletter
\renewcommand\AB@affilnote[1]{\textsuperscript{#1}\,}
\makeatother

\title{Update Opacity: Epistemic Accessibility and Governance Under AI System Change}

\author[1,2,3]{Andrea Ferrario\thanks{Corresponding author: \texttt{aferrario@ethz.ch}.} }
\affil[1]{Institute of Biomedical Ethics and History of Medicine, University of Z\"urich, Z\"urich, Switzerland}
\affil[2]{SUPSI, Dalle Molle Institute for Artificial Intelligence (IDSIA), Lugano, Switzerland}
\affil[3]{ETH Z\"urich, Z\"urich, Switzerland}

\author[4]{Joshua Hatherley}
\affil[4]{Center for Philosophy of AI, University of Copenhagen, Copenhagen, Denmark}

\begin{document}

\maketitle

\begin{abstract}
Machine learning models embedded in deployed AI systems are routinely updated to maintain correct functioning over time. Yet such updates can generate \emph{update opacity}: users may not be able to understand why the same input now yields a different output. We argue that update opacity is best understood as a \emph{diachronic failure of epistemic accessibility}: the problem is that materially relevant changes may fail to remain accessible to human users in forms that support understanding, calibrated reliance, and appropriate action under real role- and time-specific constraints. This makes update opacity a governance problem. Not all change is equally relevant, and disclosing every update would itself undermine use through overload. To address this problem, we combine two complementary governance approaches: the EU AI Act, which helps specify the system-level perimeter of normatively relevant change, and Machine Learning Operations, which provides operational tools for tracking and comparing change over time. On this basis, we propose a framework that models system change through trustworthiness profiles and trustworthiness levels, and uses threshold-based disclosure to surface materially relevant within-envelope change to different stakeholders over time. We illustrate the approach with a medical AI example and derive practical implications for lifecycle documentation, post-market monitoring, and update disclosure.
\end{abstract}


\section{Introduction}
\label{section:intro}

Machine learning (ML) models embedded in deployed artificial intelligence (AI) systems must be regularly updated to avoid performance deterioration and to maintain correct functioning over time. However, updating high-risk AI systems presents substantial risks and obstacles that have thus far received little attention in the technical or AI governance literatures \citep{ong2026considering}. One such challenge is what \citet{hatherley2025moving} refers to as the problem of update opacity, which occurs when expert users cannot, upon inspection, understand or explain why the current version of an ML model embedded in an AI system outputs $y$ on input $x$ when the previous version outputted $z\neq y$ on the same input $x$. 

The significance of update opacity is first and foremost epistemic. In fact, human use of AI systems often depends on accumulated calibration: users learn what kinds of outputs to expect, when to trust them, how to interpret warnings, and how to integrate system outputs into their own decision-making. When the system changes, this user-side understanding may no longer track the system’s current behaviour. In high-stakes domains such as healthcare, finance, or criminal justice, such misalignment can undermine appropriately grounded reliance, increase cognitive burden, and weaken downstream accountability and audit \citep{ferrario2025trustworthiness,hatherley2023diachronic,sparrow2025continuous}. 

Update opacity is therefore best understood as a problem of \emph{epistemic accessibility across diachronic AI system change in a context of use}: materially relevant change---i.e., change that is significant enough, in a context of use, to matter for justified reliance on the AI and its governance---may fail to remain accessible to users in forms they can successfully use under epistemic, practical, and institutional constraints. This form of opacity generates an AI governance problem. On the one hand, transparency about every system change is infeasible and counterproductive, as it overloads users with low-signal information and disrupts workflows. On the other hand, disclosing nothing is not acceptable, since silent behavioural change can undermine justified reliance on the AI system's outputs. Finally, halting updates \emph{tout court} is not viable either, because it invites loss of system fit in nonstationary environments. The challenge posed by update opacity is \emph{which} changes should be surfaced, \emph{when}, and \emph{to whom}. In other words: \emph{how should AI system change be made epistemically accessible over time?}

This paper addresses update opacity by combining two complementary AI governance approaches. The first is the AI trustworthiness-oriented governance regime exemplified by the EU AI Act (AIA) \citep{EU_AI_Act_2024}, which defines system-level requirements for correct functioning and identifies when change becomes normatively salient. The second is the engineering governance culture of Machine Learning Operations (MLOps) \citep{kreuzberger2023machine}, which offers practical tools for monitoring, comparing, and acting on change over time. Our claim is that neither approach is sufficient on its own. The AIA helps specify the perimeter of governance-relevant change, but it does not by itself determine how \emph{within-envelope} change should be rendered epistemically accessible to different users. MLOps, by contrast, provides operational tools for tracking change, but it does not by itself identify which monitored changes are materially relevant to justified reliance and stakeholder-facing disclosure. Our proposal combines these two approaches through a \emph{trustworthiness-based framework for governing update opacity}. We argue, first, that model-affecting changes should be treated as potential changes in the \emph{trustworthiness profile} of the AI system rather than merely as changes in model performance \citep{ferrario2025trustworthiness}. Second, we model continued correct functioning through \emph{trustworthiness levels}, which distinguish impermissible or escalation-triggering change from tolerable within-envelope change. Third, we introduce a threshold-based disclosure rule for surfacing materially relevant change to users over time. These thresholds help discriminate between background change that need not burden users and changes that are significant enough to require disclosure because they can affect calibrated reliance, oversight, or workflow expectations.

The paper is structured as follows. Section~\ref{section:update_opacity} develops the epistemic interpretation of update opacity. Section~\ref{section:AIA} discusses the governance of AI system change under the AIA. Section~\ref{section:MLOps} introduces the MLOps approach to lifecycle management and model monitoring. Section~\ref{section:our_approach} presents our trustworthiness-based framework for governing update opacity, including trustworthiness profiles, trustworthiness levels, and threshold-based disclosure. Section~\ref{section:conclusions} concludes.

\section{Update Opacity}
\label{section:update_opacity}
Modern AI systems rely on ML models to compute outputs---including predictions and recommendations, text snippets, images, and videos---that expert and lay users integrate into their decision-making processes. Key to this interaction is the recent concept of \emph{update opacity}, which Hatherley defines as follows:

\begin{quote}
Update opacity occurs when expert humans cannot, upon inspection, say why, in general, the current version of a model predicts an output ($y$) as a result of an input ($x$), \emph{while a previous version of the model predicts a different output (z), as a result of x} \citep[page~7, emphasis in original]{hatherley2025moving}.
\end{quote}

In contrast to other forms of ML model opacity \citep{guidotti2018survey,facchini2021towards,sogaard2023opacity}, update opacity is comparative and diachronic in nature:\footnote{
Hatherley distinguishes diachronic and synchronic forms of update-related variation \citep{hatherley2025moving,hatherley2023diachronic}. Diachronic variation arises when a deployed system produces different outputs at different times because of retraining, recalibration, pipeline revisions, or related changes. Synchronic variation arises when two copies of an apparently same system behave differently across sites because local data flows, update schedules, or implementation choices differ. In this paper, we focus on diachronic variation.} it concerns a discrepancy between outputs generated over time by different states of an apparently persisting AI system for the same input. As opacity should be treated as ``a concept whose meaning depends on the context of application, and on the purposes and characteristics of its users'' \citep[page~1]{facchini2021towards}, update opacity can be understood as a specifically diachronic form of \emph{access opacity}---see Section 2 in \citep{facchini2021towards}---that depends on whether the relevant user has sufficient epistemic access to update-relevant elements of the system. 

The philosophical difficulty posed by update opacity is that the user may be unable to understand and articulate \emph{why} the output changes in a way that is sufficient for justified reliance on the AI system's outputs  \citep{hatherley2025moving}. Here, we read Hatherley's phrase ``cannot, upon inspection, say why'' in epistemic terms, where to ``say why'' need not mean providing a complete causal-mechanistic explanation of the update, but, rather, the form of epistemic success that follows whenever the user has enough update-relevant understanding to identify, communicate, and appropriately act on the salient basis of the discrepancy between the earlier and the later output. 
On this interpretation, \emph{update opacity is a failure of epistemic accessibility across diachronic system change in a context of use}. It is a mismatch between an \emph{objective} phenomenon---diachronic change in the AI system and the information produced about that change---and the \emph{user-relative} cognitive and epistemic capacities of situated human agents who must consume that information successfully. Here, the \emph{context of use} includes the deployment setting of the AI system, the role-specific expectations, decision tasks, authority structures, and time pressures under which different users interact with its outputs.

There are several ways in which update opacity can arise. First, the relevant information may simply not exist in a usable form: the system may lack adequate tracking, logging, documentation, or preservation of update-relevant artifacts. Second, the information may exist but be inaccessible to the relevant user, for instance because it is buried in technical documentation, distributed across organizational units, or unavailable within the time window of decision-making. Third, the information may be available and even technically adequate, yet still fail to support epistemic uptake because it is too detailed, too abstract, too technical, or otherwise poorly matched to the user's expertise and practical constraints. This last scenario emphasizes the situated character of update opacity. Human agents using AI systems often operate under constraints of time, role, institutional authority, and workflow. A clinician, for instance, may have access to update documentation in principle, yet no realistic possibility of reading it during acute care. More generally, even where update-relevant information is available and technically adequate, epistemic success may still fail because governance and organizational constraints prevent users from seeking second opinions, escalating uncertainty, or integrating the information into ongoing decision routines. 

A further difficulty is that justified reliance on AI systems often depends on calibration that is accumulated over time. Users learn how a system behaves, when its outputs are reliable, where its weaknesses lie, and how to incorporate its outputs into their own reasoning. Yet no correct updating of this user-side understanding is guaranteed when the system itself changes. An AI system may evolve through retraining, recalibration, feature-engineering revisions, dependency updates, interface redesigns, or altered oversight procedures, while the user’s mental model remains anchored to an earlier system state. Although Hatherley formulates update opacity in terms of ML model versions, the relevant phenomenon is often produced by a broader class of \emph{model-affecting AI system updates}\footnote{These changes include modifications to data pipelines, feature processing, thresholds, dependencies, interfaces, or oversight workflows that can alter either the outputs of the embedded ML model or the way those outputs ought to be interpreted by users. In what follows, we therefore extend the discussion from \emph{ML updates} to \emph{model-affecting system updates}. Not every system update is relevant to update opacity. Only those system-level changes that can modify outputs, output distributions, or the user-facing significance of outputs in a context of use fall within its scope. More on this point in Sections~\ref{section:AIA} and~\ref{section:MLOps}.} As a result, update opacity can silently degrade the epistemic basis on which reliance had previously been justified \citep{hatherley2025moving}. In practice, this may increase cognitive burden, frustrate audit and accountability, and undermine the safe integration of AI systems into decision-making routines.\footnote{Our focus, however, is epistemic: update opacity disrupts the conditions under which users can form and maintain appropriately grounded beliefs about the significance of AI outputs over time.} This also shows why update opacity cannot be reduced to familiar problems in explainable AI. Some xAI approaches address temporality by studying how explanations, counterfactuals, or recourse procedures lose validity under model change \citep{ferrario2022robustness,barocas2020hidden,fonseca2023setting,venkatasubramanian2020philosophical}. In this body of work, however, time and model updates are typically seen as causes of the loss of validity of counterfactual explanations and algorithmic recourse procedures. Further, these approaches treat explanation primarily as a tool for local interpretability of the current model decision and for actionable guidance to affected individuals. By contrast, update opacity concerns the temporal relation between model versions and the governance of model evolution: it asks how changes across time (and across deployments) affect the user's ability to appropriately rely on the model and the AI system using it \citep{hatherley2025moving,hatherley2023diachronic}. 

For these reasons, update opacity should be understood as a problem of aligning objective diachronic system change with the conditions of human epistemic success. It requires identifying which changes are epistemically relevant to the use and oversight of the system, and making information about those changes accessible in forms that support uptake by different stakeholders under their practical constraints. We call these changes \emph{materially relevant}. Thus, materiality is neither purely technical nor purely legal: it is indexed to the practical and normative significance of change for situated users and institutions. Hatherley highlights two especially promising strategies in this space: \emph{dynamic model reporting}, which treats documentation and reporting as lifecycle practices rather than one-off disclosures, and \emph{update compatibility}, which seeks to constrain updates so that behavior remains continuous where continuity matters \citep{hatherley2025moving,bansal2019updates}. Our approach builds on these insights. More specifically, we treat update opacity as a problem of governed epistemic accessibility across system change: (1) the relevant informational perimeter of the AI lifecycle must be defined, and (2) materially relevant changes must be made accessible in ways that increase the likelihood of epistemic uptake and successful use. To show how this can be done, the next two sections introduce two complementary AI governance approaches that address different sides of this problem.

\section{The First Culture: The Governance of AI System Change in the AIA}
\label{section:AIA}

\subsection{AI Trustworthiness}
\label{subsection:AI_TW}
Over time, multiple regulations have been introducing governance schemes to manage the risks associated with updating ML models used by AI systems, especially in safety-critical domains where systems evolve after deployment. In this space, the EU Artificial Intelligence Act (AI Act) is the first horizontal regulatory framework for AI systems placed on the EU market. It adopts a principle-based structure to govern AI systems, introducing particularly demanding lifecycle obligations for \emph{high-risk AI systems}---i.e., systems whose intended purpose makes them safety- or rights-sensitive and therefore subject to stricter requirements on design, documentation, deployment, and monitoring \citep{EU_AI_Act_2024}. While other AI regulations and sectoral governance schemes are emerging globally,\footnote{Other regulations have similar mechanisms of AI governance that may be less articulated than the AIA's. In healthcare, for instance, regulators are increasingly governing machine learning updates by moving from a static approval mindset toward a lifecycle, change-control model for adaptive Software as a Medical Device (SaMD) \citep{gilbert2021algorithm}. In the United States, the FDA's framework for AI/ML-based SaMD centers on a \emph{Predetermined Change Control Plan}, combining SaMD Prespecifications (what kinds of model changes are expected/allowed) with an \emph{Algorithm Change Protocol} (methods and controls for implementing those changes while maintaining safety and effectiveness), paired with ongoing real-world performance monitoring and quality-system controls \citep{brown2021fda}.} we focus on the AIA because it is likely to exert influence on AI governance beyond the EU (via the Brussels effect \citep{Bradford2020BrusselsEffect}) and because it makes diachronic, lifecycle-oriented risk mitigation an explicit part of its regulatory architecture \citep{ferrario2026highriskidentity}. Furthermore, the AIA is binding for high-risk AI systems. However, its lifecycle governance architecture---ex ante specification of conformity-relevant requirements, post-deployment monitoring, and an identity-sensitive trigger for renewed duties under substantial modification (see below)---offers a general template for governing AI system change over time, which can be adopted as a voluntary assurance scheme for any AI system, even when not legally compulsory. For these reasons, we use the AIA as a paradigmatic, principled AI governance culture whose conceptual structure can extend beyond the high-risk perimeter.

We call the AIA governance approach \emph{principled} because its framework descends from the EU's longstanding framing of \emph{trustworthy AI}.\footnote{For instance, see Recital 7, 27, and 165 therein.}  The EU High-Level Expert Group's Ethics Guidelines for Trustworthy AI \citep{EU_HLEG_TrustworthyAI_2019} articulate trustworthiness as a lifecycle standard (lawful, ethical, and robust) and operationalize it via concrete requirements (human agency and oversight; technical robustness and safety; privacy and data governance; transparency; fairness; societal well-being; accountability). The AI Act inherits and translates this architecture into binding obligations for high-risk systems: continuous risk management, data governance for training/validation/testing data, technical documentation, logging and traceability, deployer-facing transparency and instructions for use, human oversight, and expected levels of accuracy, robustness, and cybersecurity \citep{EU_AI_Act_2024}. This package of requirements motivates a system-level notion of contextualized \emph{correct functioning}, a regulatory proxy for trustworthiness. In other words, an AI system counts as trustworthy in a context of use insofar as  it functions correctly there, namely, its functioning over time is compliant with a set of pre-defined, explicit requirements descending from AI principles.

\subsection{AI Identity Over Time and the AIA}
\label{subsection:AI_time_AIA}
The governance approach promoted by the AIA implements an implicit \emph{diachronic AI identity governance system} for high-risk AI: it specifies when an AI system continues to count as the \emph{same} regulated system over time, and when change is identity-breaking in a legal and governance-relevant sense \citep{ferrario2026highriskidentity}. This mechanism is AI lifecycle-structured. Before a high-risk system is placed on the market, providers must complete a conformity assessment (and the associated technical documentation) as a basis for CE marking, i.e., the formal declaration that the system complies with the high-risk requirements. After deployment, providers must operate a post-market monitoring system aimed at continuously collecting and analysing information about the system’s performance and risks in real-world use. Within this lifecycle regime, system updates and changes are filtered through the concept of \emph{substantial modification}: if a change is not foreseen within the originally assessed conformity envelope and it affects compliance with the high-risk requirements or alters the system's intended purpose, it is treated as no longer persisting as the same compliant artifact and renewed assessment obligations are triggered \citep{EU_AI_Act_2024}. This regulatory approach to managing changes in AI systems over time includes model-affecting system updates and is distributed across the AIA regulation (see  Table~\ref{table:EU_AI_act} for an overview of its main components).

\begin{table*}[!h]
\centering
\tiny
\setlength{\tabcolsep}{6pt}
\renewcommand{\arraystretch}{1.2}
\begin{tabularx}{\textwidth}{@{}p{3.6cm}X p{4.3cm}p{3.0cm}@{}}
\toprule
\textbf{Change type} & \textbf{Typical examples} & \textbf{Conformity-assessment relevance (high-risk AI)} & \textbf{EU AI Act reference} \\
\midrule

\textbf{Change to intended purpose} 
& Extending a clinical indication; changing target population; moving from ``decision support'' to ``automated decision;'' new deployment context that changes the system’s purpose.
& If the change modifies the \emph{intended purpose} for which the system was assessed, it is a \emph{substantial modification} and requires a \emph{new conformity assessment}. 
& Art.~3(23); Art.~43(4) \\

\textbf{Unplanned change that affects compliance with Chapter III, Section 2 requirements}
& Model-affecting system update causing materially different accuracy/robustness; changes to human oversight workflow; changes that undermine transparency/instructions; cybersecurity-relevant changes.
& If not foreseen/planned in the initial conformity assessment and it affects compliance with Section~2 requirements, it is a \emph{substantial modification} $\rightarrow$ \emph{new conformity assessment}. 
& Art.~3(23); Art.~43(4) \\

\textbf{Pre-determined changes (learning systems) assessed ex ante}
& Controlled online learning / periodic retraining where the \emph{allowed} update envelope and expected performance shifts are defined in advance.
& For systems that continue to learn: changes that were \emph{pre-determined by the provider} at the initial conformity assessment \emph{and documented} do \emph{not} count as substantial modifications (but must be captured in the technical documentation and assessed ex ante).
& Art.~43(4) (2nd para.); Annex~IV(2)(f) \\

\textbf{Software/firmware dependency and version changes}
& Upgrading runtime/OS, libraries, firmware; changes to version-update requirements; integration changes with external software components.
& Must be traceable in technical documentation. Depending on impact, such updates can become “unplanned changes affecting compliance” and therefore trigger reassessment under the substantial-modification test.
& Annex~IV(1)(c); Annex~IV(6); Art.~11(1) \\

\textbf{Data pipeline / training-validation-testing data changes}
& New data source; re-labelling; new cleaning pipeline; drift-mitigation retraining set; altered validation protocol; shifts in subgroup performance.
& Data governance is a core compliance area for high-risk AI; material data changes are conformity-relevant and, if they affect compliance, may require retraining and/or a new conformity assessment route.
& Art.~10; Annex~IV(2)(d), (2)(g); Annex~VII(4.6) \\

\textbf{Risk management and post-market monitoring driven changes}
& New risk controls; new thresholds/alerts; updated incident-handling; corrective actions after post-market evidence; updates responding to newly observed failure modes.
& These changes must be reflected in lifecycle documentation/monitoring. Post-market monitoring is specifically aimed at continuous compliance; evidence of significant risk situations may also support identifying substantial modifications.
& Annex~IV(5), (9); Art.~72(2); Art.~12(2)(a)-(b) \\

\textbf{Human oversight / transparency / UI and instructions changes}
& New interface layer; new explanation module; new override/stop procedure; updated instructions for use; revised oversight roles.
& These aspects are explicitly regulated for high-risk AI; changes may affect compliance with transparency/interpretability and oversight requirements and can therefore be conformity-relevant (and, if unplanned + compliance-affecting, substantial).
& Art.~13; Art.~14; Annex~IV(1)(g)-(h), (2)(e) \\

\bottomrule
\end{tabularx}
\caption{Main categories of changes in high-risk AI systems that are conformity-assessment relevant under the EU AI Act, including the “substantial modification” trigger for re-assessment and the special treatment of pre-determined changes for systems that continue to learn.}
\label{table:EU_AI_act}
\end{table*}

While the AIA functions as a change-control regime for socio-technical systems by anchoring governance to the continued correct functioning of high-risk AI systems over time, it cannot resolve update opacity on its own. The reason is that the AIA is designed primarily to determine when change becomes normatively salient in a regulatory sense, namely, when it affects intended purpose, compliance with high-risk requirements, or the identity of the assessed system. This remains too coarse-grained for the epistemic problem posed by update opacity.  
Furthermore, many model-affecting changes are permissible and do not trigger renewed conformity assessment, yet can still alter how users ought to understand and rely on system outputs. These model-affecting changes involve retraining, recalibration, pipeline revisions, subgroup performance shifts, dependency updates, and interface changes, which can all affect dimensions of system-level correct functioning that are central to the AIA's trustworthiness-oriented regime: transparency, human oversight, data governance, robustness, traceability, and post-market monitoring. Then, \emph{these model-affecting changes can affect dimensions that are constitutive of the AI system's trustworthiness profile}, even when they remain within the assessed trustworthiness envelope. The AIA therefore helps specify the \emph{informational perimeter} of diachronic change: it identifies \emph{which} kinds of model-affecting system updates matter for the system's continued trustworthy functioning and therefore must be documented, monitored, or escalated. What it does not yet provide is a sufficiently fine-grained account of how such change should be made epistemically accessible to different users over time, thus fully addressing update opacity. To tackle this challenge, we need to integrate the AIA with a second culture of AI governance. We turn to that culture next.

\section{The Second Culture: Managing AI System Lifecycles with MLOps}
\label{section:MLOps}
Machine learning operations (MLOps) is the domain devoted to defining and implementing operations aimed at maintaining AI systems that use ML models \citep{kreuzberger2023machine}. 
More formally, MLOps is ``a paradigm, including aspects like best practices, sets of concepts, as well as a development culture when it comes to the end-to-end conceptualization, implementation, monitoring, deployment, and scalability of machine learning products'' \citep[page~31876]{kreuzberger2023machine}.\footnote{In the Google Cloud documentation, for instance, the goal of MLOps is to ``[automate and monitor] at all steps of ML system construction, including integration, testing, releasing, deployment and infrastructure management''---see\url{https://docs.cloud.google.com/architecture/mlops-continuous-delivery-and-automation-pipelines-in-machine-learning}} \emph{Principles} or \emph{best practices} guide the different technical components of an ML model design \citep{kreuzberger2023machine}. They include CI/CD automation, orchestration, reproducibility, versioning, collaboration, continuous training and evaluation, metadata logging, continuous monitoring, and feedback loops \citep{kreuzberger2023machine}. Altogether, they enable reliable operation under nonstationary data distributions and iterative updating of ML models, procedures and architectures. 

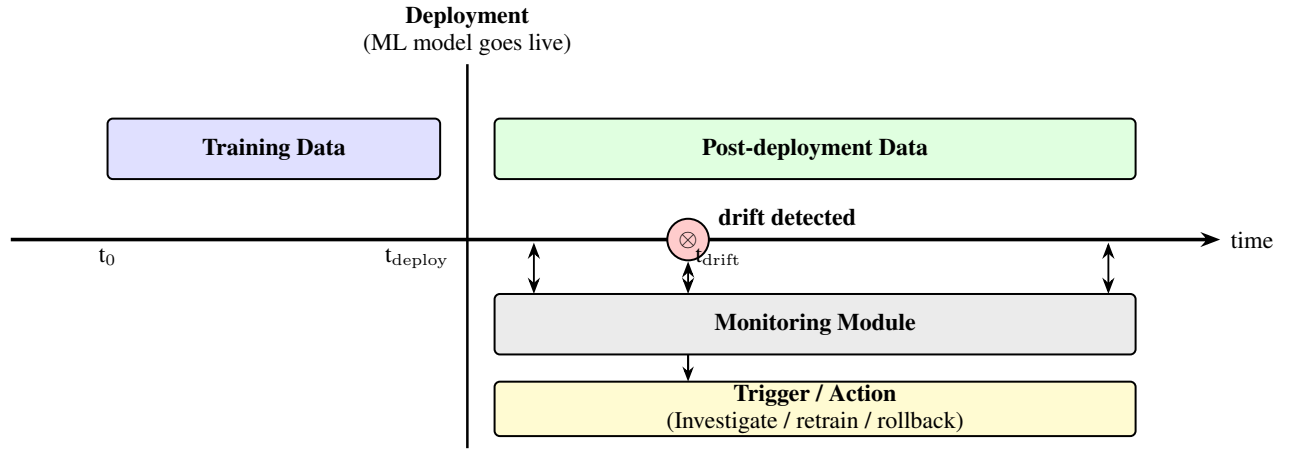
\begin{figure}[!h]
\centering
\begin{tikzpicture}[
  font=\small,
  scale=0.80,
  >=Stealth,
  axis/.style={line width=1.35pt, -{Stealth[length=2.6mm]}},
  vmark/.style={line width=0.95pt},
  callout/.style={line width=0.8pt, -{Stealth[length=2.0mm]}},
  lab/.style={font=\small},
  box/.style={rounded corners=2pt, line width=0.75pt},
  driftnode/.style={circle, draw, line width=0.75pt, fill=red!20, inner sep=1.4pt, minimum size=5.5mm}
]

\colorlet{trainfill}{blue!12}
\colorlet{postfill}{green!12}
\colorlet{monfill}{black!8}
\colorlet{actfill}{yellow!22}

\def\xL{0}
\def\xR{18}
\def\yT{0}

\def\trainL{0.6}
\def\trainR{6.1}

\def\postL{7.0}
\def\postExt{17.6}

\pgfmathsetmacro{\depX}{(\trainR+\postL)/2}

\def\dataY{1.0}
\def\boxH{1.0}

\def\monTop{-0.9}
\def\monBot{-1.9}

\def\actTop{-2.35}
\def\actBot{-3.25}

\def\driftX{10.2}

\draw[axis] (\xL-1,\yT) -- (\xR+1,\yT) node[anchor=west] {time};

\draw[box, fill=trainfill]
(\trainL,\dataY) rectangle (\trainR,\dataY+\boxH);
\node at ({(\trainL+\trainR)/2},\dataY+0.5)
{\textbf{Training Data}};

\draw[box, fill=postfill]
(\postL,\dataY) rectangle (\postExt,\dataY+\boxH);
\node at ({(\postL+\postExt)/2},\dataY+0.5)
{\textbf{Post-deployment Data}};

\draw[vmark] (\depX,\actBot-0.2) -- (\depX,\dataY+\boxH+0.9);
\node[anchor=south, align=center] at (\depX,\dataY+\boxH+0.9)
{\textbf{Deployment}\\(ML model goes live)};

\node[driftnode] (driftsym) at (\driftX,\yT) {$\otimes$};
\node[lab, anchor=west] at (\driftX+0.35,\yT+0.38) {\textbf{drift detected}};

\draw[box, fill=monfill]
(\postL,\monBot) rectangle (\postExt,\monTop);
\node at ({(\postL+\postExt)/2},{(\monTop+\monBot)/2})
{\textbf{Monitoring Module}};

\draw[<->, line width=0.7pt]
  (\postL+0.65,\yT-0.05) -- (\postL+0.65,\monTop);

\draw[<->, line width=0.7pt]
  (driftsym.south) -- (\driftX,\monTop);

\draw[<->, line width=0.7pt]
  (\postExt-0.45,\yT-0.05) -- (\postExt-0.45,\monTop);

\draw[box, fill=actfill]
(\postL,\actBot) rectangle (\postExt,\actTop);
\node[align=center] at ({(\postL+\postExt)/2},{(\actTop+\actBot)/2})
{\textbf{Trigger / Action}\\(Investigate / retrain / rollback)};

\draw[callout] (\driftX,\monBot) -- (\driftX,\actTop);

\node[anchor=north] at (\trainL,\yT) {t$_0$};
\node[anchor=north east] at (\depX-0.15,\yT) {t$_{\mathrm{deploy}}$};
\node[anchor=north] at (\driftX+0.5,\yT) {t$_{\mathrm{drift}}$};
\end{tikzpicture}
\caption{Drift detection and update triggering in MLOps: after deployment, a monitoring module tracks post-deployment data over time; when drift is detected (marked by $\otimes$ at $t_{\mathrm{drift}}$), the monitoring signal triggers operational actions such as investigation, retraining, or rollback.}
\label{fig:mlops_drift_timeline}
\end{figure}

The \emph{technical components} of an AI system lifecycle under the MLOps paradigm include CI/CD, source code repository, workflow orchestration, feature store system, model training infrastructure and registry, ML model metadata stores, model serving and monitoring \citep{kreuzberger2023machine}. In particular, the \emph{model monitoring component} typically covers (1) model performance tracking (e.g., predictive quality and calibration), (2) data quality and drift,  and (3) infrastructure and orchestration quality \citep{kreuzberger2023machine,eken2025multivocal,bayram2024adaptive,lu2018learning}.
Ultimately, the model monitoring component can trigger interventions ranging from alerts and additional validation checks to model update, including rollback or removal from production, depending on the severity of the detected change and the system's operating constraints.
A notable trigger for model update under the MLOps paradigm is the \emph{detection of drift.} Intuitively, drift denotes a statistical change in the data-generating process after model deployment, such that the distribution of operational inputs (or the input-output relation) departs from the baseline conditions under which the model was developed.\footnote{There are different forms of drift, including covariate, label, and concept drift. More details in \citep{liu2013change,liu2022concept}.} The idea behind drift detection is to compare time-indexed operational data to reference data---often the training data of the model currently deployed and under monitoring---to assess whether newer inputs remain qualitatively and quantitatively similar to those on which the model was developed. In practice, such monitoring is implemented as a continuous pipeline: time-series input data are collected, quality assessments are run, drift indicators are computed, and threshold-based decision rules determine whether additional scrutiny or corrective actions (e.g., retraining) are required \citep{bayram2024adaptive}. Figure~\ref{fig:mlops_drift_timeline} shows the general schema of drift detection.

A plethora of drift indicators and decision rules are currently considered in applications. For instance, many drift monitoring procedures rely on divergence-based measures computed on marginal (or low-dimensional) distributions, complemented by data-quality checks and statistical tests \citep{liu2013change,liu2022concept,bayram2024adaptive}. In this setup, the training distribution plays the role of a baseline that the operational distribution is expected to approximate well in the absence of high levels of drift. A general, key design choice is then the drift decision rule, which is usually threshold-based: drift is flagged when one or more divergences exceed a predefined threshold or as a result of statistical tests, triggering pre-emptive mitigation actions that include model updates; otherwise, no action is taken \citep{liu2013change,bayram2024adaptive}. In nonstationary environments, however, fixed thresholds are often inadequate, motivating adaptive thresholds that track baseline variability and update the alarm criterion over time \citep{bayram2024adaptive,liu2013change}. The development of context-based, time-adaptive thresholds to detect drift is an open avenue of research.

MLOps is therefore a principled governance culture rooted in engineering best practices. It operationalises norms such as reliability, reproducibility, traceability, and safe iteration by instrumenting systems with continuous telemetry, comparing time-indexed behaviour to baselines, and converting continuous measurements into governance-relevant events through thresholded trigger--action playbooks. Yet MLOps is not sufficient to address update opacity. The reason is that MLOps is oriented toward operational control: it provides measurement, versioning, monitoring, and intervention triggers, but it does not by itself determine which monitored changes are \emph{epistemically material} for justified reliance, oversight, or stakeholder-facing disclosure. Furthermore, as noted in Section~\ref{section:AIA}, update opacity is driven by model-affecting system updates: those system-level interventions that can modify outputs, output distributions, or the practical meaning of outputs for users. Even when these changes are tracked internally by MLOps procedures, the resulting artifacts are typically optimized for engineers rather than for deployers or other users. Standard MLOps deliverables do not, by themselves, tell a clinician, loan officer, or risk manager what has changed in a form that supports timely epistemic uptake. These stakeholders would not benefit from being told that yesterday's output would have differed because an overnight retraining shifted the weights of a deep learning model, or because a revised cross-validation procedure corrected data leakage introduced by an earlier feature-normalization bug. In the terms of Section~\ref{section:update_opacity}, MLOps supplies the operational machinery needed to \emph{track} diachronic change, but not yet a principled account of which changes should be communicated to which users, and in what form, in order to preserve their calibrated reliance over time.

\vspace{1em}

In summary, the two governance cultures presented in Sections~\ref{section:AIA} and~ \ref{section:MLOps} are complementary rather than competing. The AIA supplies the normative machinery that determines which kinds of change matter for continued system-level correct functioning and compliance over time. MLOps supplies the operational machinery for tracking, comparing, and acting on change. \emph{Update opacity emerges in the shared permissibility space between them}: MLOps detects and often motivates updates within a system that still remains, in AIA terms, within its conformity-assessed envelope. In that space, the system may remain formally deployable while becoming materially less legible to users. This is the starting point of our approach.

\subsection{A Clarification: AI Reliability vs. Trustworthiness}
\label{subsection:rel_vs_tw}

A brief clarification of the terms \emph{reliability} and \emph{trustworthiness} is useful before introducing our approach. Reliability is commonly understood as the continued correct operation of a system under expected conditions and over time \citep{NIST_AIRMF_2023}.\footnote{Similarly, ISO/IEC TS 5723:2022 defines reliability as ``the ability of an item to perform as required, without failure, for a given time interval, under given conditions.''} In this sense, reliability captures an important part of what is at stake in governing AI system change: whether the system continues to perform as intended in its context of use. Trustworthiness, however, is broader. In the European policy tradition and related standards, trustworthy AI includes not only reliable performance, but also dimensions such as transparency, accountability, privacy, data governance, robustness, human oversight, and related socio-technical requirements \citep{EU_HLEG_TrustworthyAI_2019,OECD_TrustworthyAI_2021}. Thus, trustworthiness includes correct functioning in the sense captured by reliability, but is not exhausted by it.\footnote{Similarly, the OECD frames trustworthy AI in terms of systems that are robust, secure, safe,  fair, transparent, explainable, and accountable \citep{OECD_TrustworthyAI_2021}. Relatedly, ISO/IEC TS 5723:2022 defines AI trustworthiness as the ``ability [of an AI system] to meet stakeholders' expectations in a verifiable way.'' These standards explicitly include accountability, accuracy, availability, controllability, integrity, privacy, \emph{and reliability}, among other properties of AI trustworthiness.}  The two constructs overlap around the idea of correct functioning of AI systems over time, while trustworthiness expands that engineering core into a broader socio-technical and normative governance profile. 

\section{Merging the Two Cultures: Our Approach}
\label{section:our_approach}

\subsection{Our Approach: Core Idea}
If update opacity is a failure of epistemic accessibility across system change in a context of use, then a satisfactory response must  identify which changes are materially relevant to the system's continued trustworthy functioning and render those changes accessible to human users in forms that support epistemic uptake over time. This is the core idea of our approach.

The crucial point is that not all model-affecting system updates matter equally to address update opacity. Some are so significant that they alter intended purpose, compliance status, or the conditions under which the system continues to count as the same admissible artifact. Others remain within the assessed trustworthiness envelope, yet still become important for justified reliance, oversight, or workflow because they change outputs, subgroup behaviour, or the practical meaning of outputs for users. Our framework is designed for this second class of cases. This is why a framing focused on model performance is insufficient. Model-affecting system updates can alter not only predictive validity, but also robustness under shift, data governance in operation, traceability and logging, cybersecurity posture, the effectiveness of human oversight, and transparency to deployers and other stakeholders. Once update opacity is seen as a system-level phenomenon of change in context, the appropriate governance object is not the model alone, but the AI system's \emph{trustworthiness profile}, namely, the structured set of requirements that define its correct functioning over time \citep{ferrario2025trustworthiness}.

Our approach elaborates this idea in three steps:

\begin{enumerate}[leftmargin=*, noitemsep]
\item \textbf{discriminating between tolerable and non-tolerable change within the system's conformity envelope (AIA)},
\item \textbf{introducing the trustworthiness profile and  the trustworthiness level function of an AI system (AIA+MLOps)}, 
\item \textbf{defining thresholded trigger-action governance to disclose materially relevant model-affecting system updates (MLOps)}.
\end{enumerate}

Our approach rules out disclosing all changes, which collapses transparency into noise, and  disclosing nothing, which undermines appropriate reliance and accountability. It also rules out halting ML model updates altogether, which is incompatible with maintaining the correct functioning of AI systems in nonstationary environments. It follows a proportionality criterion instead: it triggers disclosure when within-envelope change becomes materially relevant in a context of use, even if the system continues to function as expected in the regulatory sense. Let us discuss the three steps of our approach in some detail.

\subsection{The Three Steps of Our Approach}
\paragraph{Step 1: Tolerable vs. non-tolerable AI system changes.}
We distinguish two categories of changes that matter for governing update opacity under the AIA. We call the \emph{assessed trustworthiness envelope} of an AI system the set of foreseen and admissible changes under which the system continues to satisfy the trustworthiness conditions specified \emph{ex ante} and therefore does not trigger renewed conformity duties. This envelope is the overall space of foreseen and governance-admissible change. Then, \textit{tolerable changes} are those model-affecting system updates that remain within this assessed trustworthiness envelope: they do not violate the diachronic identity of the deployed system, alter its intended purpose, or affect compliance with high-risk requirements in a way that was not foreseen and controlled within the conformity assessment. Under such changes, the system remains within its assessed trustworthiness envelope and retains its governance-admissible status. \textit{Non-tolerable changes}, by contrast, are those changes that fall outside this envelope and therefore trigger renewed conformity duties, reassessment, rollback, or other forms of escalation. They mark a break in the system’s admissible diachronic persistence, as discussed in Section~\ref{section:AIA}.
\paragraph{Step 2: AI trustworthiness profiles and functions.}
Let $A$ denote an AI system. Its \textbf{trustworthiness profile} is the set
\[
\mathcal{P}^A=\{(d^A_1,m^A_1),\dots, (d^A_n,m^A_n)\},
\]
where each $d^A_i$ is a predetermined, quantifiable, and time-relative trustworthiness dimension of $A$, and $m^A_i$ is the corresponding measurement protocol, for all $i=1,\dots,n$ \citep{ferrario2025trustworthiness,ferrario2024justifying}. The profile $\mathcal{P}^A$ is defined \emph{ex ante} as part of the system’s conformity assessment and specifies what it means for $A$ to function correctly in its intended context of use.\footnote{Here, we formalise the \emph{actual} trustworthiness of system $A$ over time, that is, the extent to which the artefact is capable of functioning correctly throughout its lifecycle. This is distinct from \emph{perceived} trustworthiness, namely, an individual’s time-relative subjective estimate of how trustworthy the system is. For further discussion of this distinction, see \citet{ferrario2024justifying}.} Depending on the system and domain, dimensions may include predictive performance, robustness under shift, data-pipeline integrity, safety and cybersecurity, and the effectiveness of human oversight. 

Each dimension can be specified at different levels of granularity; providers and deployers should specify an appropriate level of specification given the intended purpose and operating conditions. Each dimension in $\mathcal{P}^A$ must be operationalised by one or more measurement protocols. Performance may be tracked via F1 score or mean squared error, robustness via drift indicators, subgroup stability, or stress tests. Multiple indicators for the same dimension may be aggregated into composite scores \citep{ferrario2024justifying}. There is, however, no unique method that fixes which metrics, thresholds, or scoring functions are appropriate. These choices are context-sensitive, depending on the intended purpose and operating constraints of the system, and must be justified on a case-by-case basis. For that reason, the selected protocols (and their rationales) should be explicitly documented and maintained in the system's conformity-assessment documentation as part of the trustworthiness profile. Recent work on AI-agent reliability offers a useful illustration of this general strategy: \citet{rabanser2026towards} distinguish dimensions such as consistency, robustness, predictability, and safety, and operationalise them through multiple quantitative measures---see Table 1 in \citep{rabanser2026towards}. Our framework generalises this idea from reliability assessment to the broader governance of AI trustworthiness under update.

At any time $t$, we realise the profile $\mathcal{P}^A$ by collecting the measurements prescribed by its protocols into a vector
\[
\mathbf{m}^A:[t_0,\infty)\to\mathbb{R}^n,\qquad
t\mapsto \mathbf{m}^A(t)=\big(m^A_1(t),\dots,m^A_n(t)\big),
\]

which provides a time-indexed quantification of the trustworthiness profile of $A$. Next, let $L^A:\mathbb{R}^n\to\{1,\dots,K\}$ be a level assignment function that partitions the measurement space into $K$ discrete trustworthiness levels.
We then define the (time-indexed) \textbf{trustworthiness level function} of $A$ \citep{ferrario2025trustworthiness} as the composite map

\begin{equation}
\texttt{T}^A:[t_0,\infty)\to\{1,\dots,K\},
\qquad
\text{where}\qquad
\texttt{T}^A := L^A\circ \mathbf{m}^A,
\qquad\text{so that}\qquad
\texttt{T}^A(t)=L^A\!\big(\mathbf{m}^A(t)\big).
\label{eq:TW_levels}
\end{equation}

\begin{figure}
    \centering
\begin{tikzpicture}
\begin{axis}[
  width=8cm,
  height=5.5cm,
  xmin=0, xmax=1,
  ymin=0.5, ymax=3.5,
  xlabel={Accuracy},
  ylabel={Trustworthiness Level},
  xtick={0,0.68,0.9,1.0},
  ytick={1,2,3},
  axis lines=left,
  tick align=outside,
  enlargelimits=false,
]

\addplot[line width=1pt] coordinates {(0.00,1) (0.68,1)};
\addplot[line width=1pt] coordinates {(0.68,2) (0.90,2)};
\addplot[line width=1pt] coordinates {(0.90,3) (1.00,3)};

\addplot[
  only marks,
  mark=*,
  mark size=1.5pt
] coordinates {
  (0.68,1)
  (0.90,2)
  (1.00,3)
};

\addplot[
  domain=0:1,
  samples=300,
  smooth,
  line width=0.4pt
] ({x}, {1 + 1/(1 + exp(-35*(x-0.68))) + 1/(1 + exp(-95*(x-0.90)))});

\end{axis}
\end{tikzpicture}
    \caption{Toy example of a trustworthiness level function. The trustworthiness profile has one dimension (accuracy). The step-wise mapping yields discrete plateaux, while the curve illustrates a smooth alternative that is sensitive to small accuracy fluctuations near the transition points $0.68$ and $0.9$, potentially increasing governance strain.}
    \label{fig:tw_1d}
\end{figure}
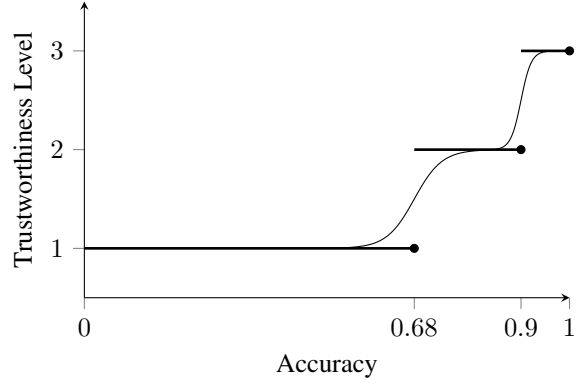

At any time $t$, $\texttt{T}^A(t)$ represents the trustworthiness level of the system $A$ at that time. A few AI governance considerations motivate representing trustworthiness levels via step-wise mappings such as \eqref{eq:TW_levels}. Consider the toy example in Fig.~\ref{fig:tw_1d}, where the trustworthiness profile consists of a single dimension, accuracy, measured on $[0,1]$, and where three trustworthiness levels correspond to a partition of that interval. By design, the system remains in the highest level even if accuracy decreases slightly from above $0.9$; however, crossing $0.9$ induces a \emph{level jump}, signalling a loss of operational fit with respect to the predefined conformity requirements. By contrast, a smooth mapping from accuracy to levels---see the curve in Fig.~\ref{fig:tw_1d}---tends to introduce brittleness: small fluctuations around transition points may repeatedly change the assessed level, thereby inducing governance strain---e.g., by recurrent escalation or reassessment---even when the system remains within an acceptable operating regime.

More generally, a step-wise representation allows updates that move the system \emph{within} a plateau---as $\mathbf{m}^A(t)$ varies without changing level---while treating boundary crossings as governance-relevant events. This reflects how lifecycle assurance regimes are typically organised around discrete decision states (e.g., continued deployability vs.\ escalation) and motivates modelling trustworthiness as a finite set of regimes (our levels). Importantly, the plateau view captures two complementary facts: (1) many AI system changes are tolerable (no change of level), and (2) within-level change can still become materially relevant for calibrated reliance and safeguards. We address this second point below in Step~3. 
Finally, the discrete structure of $\texttt{T}^A$ can be implemented using interpretable models: given historical measurements of trustworthiness dimensions and qualitative level judgments, one can learn $L^A$ (and hence $\texttt{T}^A$) using decision trees or rule lists. This makes it possible to identify which dimensions drive level assignments or transitions, thereby supporting reporting and  transparency.

\paragraph{Step 3: Thresholded trigger-action governance to disclose model-affecting system updates.}
As discussed above, model-affecting system updates are frequent in many deployments and can be triggered by a variety of causes, including drift, pipeline revisions, recalibration, dependency changes, and related lifecycle events. Importantly, many such updates remain \emph{within} the system's assessed trustworthiness envelope: they do not induce a trustworthiness level \emph{jump}, and the system stays on the same admissible plateau. Thus, governance responses to update opacity must focus especially on within-level change that can remain formally admissible while becoming epistemically material for users in a context of use. This is the goal of this last step of our approach. Drawing on the MLOps governance culture, we introduce a thresholded rule to discriminate materially relevant from immaterial model-affecting system updates.

Let us suppose that system $A$ is deployed at time $t_0$ with trustworthiness level $\texttt{T}^A(t_0)$. Let $t_1<t_2<\dots<t_n$ denote times at which a model-affecting system update inside $A$ is released. At each $t_i$, we compute the corresponding trustworthiness level $\texttt{T}^A(t_i)=L^A(\mathbf{m}^A(t_i))$. Assume that no level jump occurs over $[t_0,t_n]$; that is,
all updates remain on the same trustworthiness plateau.\footnote{Mathematically: $\texttt{T}^A(t_1)=\texttt{T}^A(t_2)=\cdots=\texttt{T}^A(t_n)=\texttt{T}^A(t_0)$. If a level jump occurs, the system exits its admissible regime and governance should escalate accordingly (e.g., corrective action, rollback, or renewed conformity duties, depending on the assurance regime).} To detect within-plateau changes that are large enough to plausibly disrupt appropriate reliance and, therefore, require transparency measures, we introduce a baseline-relative \textbf{dissimilarity} function
\[
\Delta^A:\mathbb{R}^n\times \mathbb{R}^n \to \mathbb{R}_{\ge 0},
\qquad
(\mathbf{m}^A(t_i),\mathbf{m}^A(t_j))\mapsto \Delta^A(\mathbf{m}^A(t_i),\mathbf{m}^A(t_j)),
\]

which quantifies how much the realised profile $\mathbf{m}^A(t_j)$ departs from a reference baseline $\mathbf{m}^A(t_i)$, with $t_j\geq t_i$.\footnote{It is customary to assume that $\Delta^A(\mathbf{m}^A(t_i),\mathbf{m}^A(t_i))=0$. Further, unlike probabilistic divergences, $\Delta^A$ is not restricted to probability distributions. It can be defined over heterogeneous measurement vectors by combining dimension-specific normalisations and weights, and it can be chosen to reflect governance priorities (e.g., penalising degradations more than improvements).}  We then define the \textbf{update opacity transparency trigger rule} as follows:

\begin{equation}
\texttt{IF}\ \ \Delta^A\!\big(\mathbf{m}^A(t_0), \mathbf{m}^A(t_i)\big)>\epsilon^A
\ \ \texttt{THEN}\ \ \text{Disclose model-affecting system update.}
\label{eq:disclosure_rule}
\end{equation}

Here, $\epsilon^A>0$ is a context-relative threshold (hence the superscript $A$) chosen \emph{strictly inside} the admissible region defined by the plateau. 

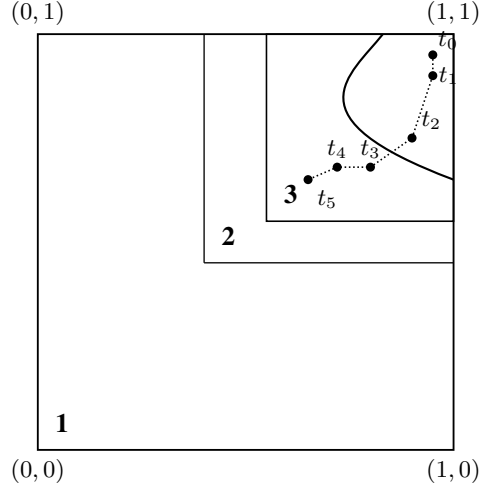
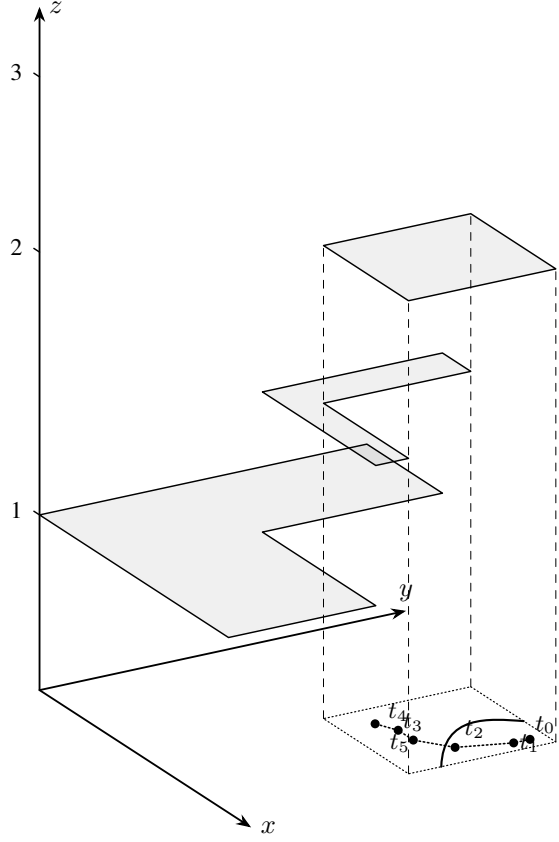
\begin{figure}[!h]
\centering

\begin{subfigure}[t]{0.48\textwidth}
\centering
\begin{tikzpicture}[
  scale=5.5,
  frame/.style={line width=0.7pt},
  gridline/.style={line width=0.5pt},
  region/.style={line width=0.6pt},
  traj/.style={line width=0.6pt, densely dotted},
  pt/.style={circle, fill, inner sep=1.2pt},
  lab/.style={font=\small}
]

\draw[frame] (0,0) rectangle (1,1);

\node[lab, anchor=north]  at (0,0) {$(0,0)$};
\node[lab, anchor=north]  at (1,0) {$(1,0)$};
\node[lab, anchor=south] at (0,1) {$(0,1)$};
\node[lab, anchor=south]  at (1,1) {$(1,1)$};

\draw[region] (0.55,0.55) rectangle (1,1);

\draw[gridline] (0.4,1) -- (0.4,0.45);
\draw[gridline] (0.4,0.45) -- (1,0.45);

\coordinate (t0) at (0.95,0.95);
\coordinate (t1) at (0.95,0.90);
\coordinate (t2) at (0.90,0.75);
\coordinate (t3) at (0.80,0.68);
\coordinate (t4) at (0.72,0.68);
\coordinate (t5) at (0.65,0.65);

\node[pt] at (t0) {};
\node[lab, xshift=-1.5pt, yshift=-1.25pt, anchor=south west] at (t0) {$t_0$};

\node[pt] at (t1) {};
\node[lab, xshift=-1.25pt, anchor=west] at (t1) {$t_1$};

\node[pt] at (t2) {}; \node[lab, anchor=south west] at (t2) {$t_2$};
\node[pt] at (t3) {}; \node[lab, anchor=south]      at (t3) {$t_3$};
\node[pt] at (t4) {}; \node[lab, anchor=south]      at (t4) {$t_4$};
\node[pt] at (t5) {};
\node[lab, yshift=-0.5pt, anchor=north west] at (t5) {$t_5$};

\draw[traj] (t0)--(t1)--(t2)--(t3)--(t4)--(t5);

\node[anchor=south west] at (0.02,0.02) {\textbf{1}};
\node[anchor=south west] at (0.42,0.47) {\textbf{2}};
\node[anchor=south west] at (0.57,0.57) {\textbf{3}};

\draw[line width=0.8pt]
  (0.83,1.00)
    .. controls (0.75,0.9) and (0.6,0.8) .. (1.00,0.65);

\end{tikzpicture}
\caption{2D view (top-down).}
\label{fig:tw_2d_panel}
\end{subfigure}
\hfill

\begin{subfigure}[t]{0.48\textwidth}
\centering
\tdplotsetmaincoords{68}{60}
\begin{tikzpicture}[tdplot_main_coords, scale=5, line join=round, line cap=round]

\tikzset{
  traj/.style={line width=0.6pt, densely dotted},
  pt/.style={circle, fill, inner sep=1.2pt},
  lab/.style={font=\small},
  lvl/.style={font=\small},
  surf/.style={draw, line width=0.55pt, fill opacity=0.06},
  axis/.style={-{Stealth[length=2.0mm]}, line width=0.7pt},
  vconn/.style={line width=0.20pt, dashed},
  baserect/.style={line width=0.45pt, densely dotted}
}

\def\zOne{1}
\def\zTwo{1.75}
\def\zThr{2.25}
\def\zBase{0.5}

\coordinate (L1A) at (0,0,\zOne);
\coordinate (L1B) at (1,0,\zOne);
\coordinate (L1C) at (1,0.45,\zOne);
\coordinate (L1D) at (0.4,0.45,\zOne);
\coordinate (L1E) at (0.4,1,\zOne);
\coordinate (L1F) at (0,1,\zOne);

\coordinate (L2A) at (0.4,1,\zTwo-0.35);
\coordinate (L2B) at (0.4,0.45,\zTwo-0.35);
\coordinate (L2C) at (1,0.45,\zTwo-0.35);
\coordinate (L2D) at (1,0.55,\zTwo-0.35);
\coordinate (L2E) at (0.55,0.55,\zTwo-0.35);
\coordinate (L2F) at (0.55,1,\zTwo-0.35);

\coordinate (L3A) at (0.55,0.55,\zThr-0.4);
\coordinate (L3B) at (0.55,1,\zThr-0.4);
\coordinate (L3C) at (1,1,\zThr-0.4);
\coordinate (L3D) at (1,0.55,\zThr-0.4);

\filldraw[surf] (L1A)--(L1B)--(L1C)--(L1D)--(L1E)--(L1F)--cycle;
\filldraw[surf] (L2A)--(L2B)--(L2C)--(L2D)--(L2E)--(L2F)--cycle;
\filldraw[surf] (L3A)--(L3B)--(L3C)--(L3D)--cycle;

\coordinate (B0A) at (0.55,0.55,\zBase);
\coordinate (B0B) at (0.55,1,\zBase);
\coordinate (B0C) at (1,1,\zBase);
\coordinate (B0D) at (1,0.55,\zBase);

\draw[baserect] (B0B)--(B0C)--(B0D)--(B0A)--cycle;

\draw[vconn] (B0A) -- (L3A);
\draw[vconn] (B0B) -- (L3B);
\draw[vconn] (B0C) -- (L3C);
\draw[vconn] (B0D) -- (L3D);

\coordinate (t0) at (0.95,0.95,\zBase);
\coordinate (t1) at (0.95,0.90,\zBase);
\coordinate (t2) at (0.90,0.75,\zBase);
\coordinate (t3) at (0.80,0.68,\zBase);
\coordinate (t4) at (0.72,0.68,\zBase);
\coordinate (t5) at (0.65,0.65,\zBase);

\node[pt] at (t0) {};
\node[lab, xshift=-1.5pt, yshift=-1.25pt, anchor=south west] at (t0) {$t_0$};

\node[pt] at (t1) {};
\node[lab, xshift=-1.25pt, anchor=west] at (t1) {$t_1$};

\node[pt] at (t2) {};
\node[lab, anchor=south west] at (t2) {$t_2$};

\node[pt] at (t3) {};
\node[lab, anchor=south] at (t3) {$t_3$};

\node[pt] at (t4) {};
\node[lab, anchor=south] at (t4) {$t_4$};

\node[pt] at (t5) {};
\node[lab, yshift=-0.5pt, xshift=2pt, anchor=north west] at (t5) {$t_5$};

\draw[traj] (t0)--(t1)--(t2)--(t3)--(t4)--(t5);

\draw[line width=0.8pt]
  (0.83,1.00,\zBase)
    .. controls (0.77,0.93,\zBase) and (0.68,0.84,\zBase) .. (1.00,0.65,\zBase);

\draw[axis] (0,0,\zBase) -- (1.12,0,\zBase) node[anchor=west] {$x$};
\draw[axis] (0,0,\zBase) -- (0,1.12,\zBase) node[anchor=south] {$y$};
\draw[axis] (0,0,\zBase) -- (0,0,2.45) node[anchor=west] {$z$};

\foreach \zz/\lbl in {\zOne/1,\zTwo/2,\zThr/3}{
  \draw[line width=0.55pt] (0,0,\zz) -- (-0.03,0,\zz);
  \node[lvl, anchor=east] at (-0.04,0,\zz) {\lbl};
}

\end{tikzpicture}
\caption{3D view (stepped plateaux).}
\label{fig:tw_3d_panel}
\end{subfigure}
\caption{Trustworthiness plateaux and update disclosure in 2D and 3D. Panel (a) shows the two-dimensional trustworthiness profile with three trustworthiness levels and an update trajectory $t_0,\ldots,t_5$ on the highest plateau. Panel (b) visualizes the same construction in three dimensions, with discrete trustworthiness levels along the $z$-axis and the level-3 footprint projected onto two dimensions.}
\label{fig:tw_2d_3d_panels}
\end{figure}

The intuition is familiar from engineering: a machine may remain operational, yet an increasing vibration index signals emerging abnormal behaviour. In this sense, $\epsilon^A$ is an \emph{epistemic materiality threshold} for within-envelope change: it converts an admissible update into a governance-relevant transparency event when the update becomes large enough to plausibly disrupt appropriate reliance, workflow expectations, or accountability practices. In practice, $\Delta^A$ can be instantiated using the same monitoring summaries that MLOps teams already compute to compare a newly trained ML model candidate against a production baseline. A simple choice is an $\ell_1$ distance $\Delta^A\!\big(\mathbf{m}^A(t_0), \mathbf{m}^A(t_i)\big)=\|\mathbf{m}^A(t_0)- \mathbf{m}^A(t_i)\|_1$. This $\Delta^A$ specification can become asymmetric, for instance, by weighting realizations $\mathbf{m}^A(t)$ over time. Figure~\ref{fig:tw_2d_3d_panels} illustrates the mechanism behind rule~\eqref{eq:disclosure_rule}: the system remains on the highest plateau (level~3), but successive updates trace a trajectory on that plateau. The threshold boundary separates updates that remain sufficiently close to the deployment baseline ($t_0$) from those that are far enough to require disclosure under \eqref{eq:disclosure_rule}. In the example, updates at $t_1$ and $t_2$ do not trigger disclosure, whereas $t_3$, $t_4$, and $t_5$ do.

In summary, \emph{our approach treats update opacity as an epistemic accessibility problem addressed by an AI trustworthiness governance framework}. We control for model-affecting system updates by dynamically tracking their effects on the different dimensions of AI trustworthiness. The key benefit of our approach is practical: rather than demanding full transparency about every possible update---an infeasible task---providers can maintain transparency by issuing warnings or disclosures when within-envelope change becomes materially relevant for deployers and affected parties.


\subsection{A Medical AI Example: Stroke Triage Decision Support}
\label{subsection:example_stroke}
Let us consider a medical AI system used in emergency stroke triage. The system ingests multimodal inputs---for instance, non-contrast CT scans, CT perfusion maps, clinical variables such as stroke score, symptom onset time, and patient age---and returns a structured recommendation that must be acted on within a narrow time window: treat here, transfer to a comprehensive stroke centre, or monitor and re-evaluate. The stakes of this output are high: a misclassification in either direction---undertriaging a large vessel occlusion, or overtriaging a mimic---carries the risk of direct patient harm or resource consequences. In this setting, physicians do not have the time to interrogate an ML model, reconstruct a pipeline history, or read lengthy technical documentation \citep{sparrow2020high}. Instead, safe use depends on accumulated calibration: they learn how the system behaves, when its outputs are reliable, and how its warnings fit into existing decision routines. This tacit knowledge, built through repeated interaction, is a key mechanism through which the system is safely integrated into the clinical workflow. This is precisely why update opacity is so disruptive in acute care. When the system changes, even slightly, the practical understanding built through repeated interaction may no longer track the current system state.


Suppose that the deployed system has been in use for eight months and remains within the highest trustworthiness level. Its aggregate discrimination on a prospective validation cohort remains above the conformity-assessed threshold, its calibration curve is within tolerance, and no substantial modification has been triggered under the AIA. Yet, as part of ordinary MLOps practice, the provider pushes a model update in response to covariate drift caused by a firmware upgrade on the CT scanners across the hospital network. The update retrains the model on a rolling window of newer cases and recalibrates the decision threshold to correct a marginal shift in false-positive rate. Aggregate performance (e.g., AUROC) improves slightly. From a purely global perspective, the update appears beneficial and remains well within the admissible envelope: no changes in trustworthiness level occur. 

However, the improvement conceals behaviourally important differences. Because the post-upgrade cases are over-represented in the retraining data, the relative weight of perfusion-map features shifts. As a result, transfer recommendations are now generated more frequently for some elderly patients with moderate stroke scores; borderline cases near the treat/transfer boundary flip recommendation more often than before; and overnight alert patterns change in ways that experienced staff notice almost immediately. None of these changes, taken in isolation, triggers a level jump. Taken together, they nevertheless alter the conditions under which clinicians had learned to rely on the system. This is a paradigmatic instance of update opacity within continued admissibility.

Our framework responds to this situation by treating the update as potentially material even though it remains within the admissible trustworthiness plateau. Once the dissimilarity threshold is crossed, the update becomes disclosure-relevant. This disclosure should not take the form of a generic changelog entry or a PDF added to a repository. In acute stroke triage, update-relevant information must be surfaced within the clinical workflow itself. For instance, a minimal banner in the triage interface might state: ``Model updated [date]. Transfer recommendations for patients over 75 may differ from previous behaviour. Tap for details and comparisons.'' A one-click summary can then provide the decision-relevant delta: which recommendation categories changed, for which subgroups, in which direction, and with what expected workflow implications. In parallel, a fuller audit-grade log can record update metadata, validation deltas, and conformity documentation for the relevant governance leads.

This example illustrates two general points. First, update opacity is not simply a technical issue about drift or performance maintenance. It is a problem about preserving the epistemic conditions of appropriate reliance as AI systems evolve. Second, disclosure must be role-sensitive. Frontline clinicians need concise, actionable information embedded in the point of use; governance leads and auditors need richer documentation that supports post-market monitoring and accountability. The same logic applies beyond medicine, including radiology decision support, sepsis early-warning systems, credit scoring tools, and recidivism assessment systems. In each case, the core task is the same: to make materially relevant system change epistemically accessible over time without collapsing transparency into overload. Our approach offers
a principled, operationalisable answer to this problem.


This example also illustrates why update opacity has direct accountability consequences that extend beyond clinical safety. Consider a scenario in which a patient is undertriaged---that is, sent to monitor and re-evaluate when they should be transferred to comprehensive care---and subsequently deteriorates. In the later review, it emerges that the triage decision was made four days after thefour days after the model-affecting system update, by a registrar who was unaware that the system's behaviour for elderly patients had shifted. The provider can demonstrate that the system remained within its conformity envelope throughout. The deploying hospital can demonstrate that its staff were trained on the system at deployment. However, neither party can demonstrate that the registrar had adequate, actionable knowledge of the update-induced behaviour change at the time of the decision. This is precisely the accountability lacuna that update opacity creates, and that our disclosure mechanism is designed to close. By triggering a documented, workflow-embedded disclosure at the point where the dissimilarity threshold is crossed, the framework establishes a clear chain of epistemic responsibility: the provider discloses, the deployer receives and acknowledges, and the frontline user is informed in a manner proportionate to their role and time constraints. If a disclosure is triggered and acknowledged, subsequent decisions are made under conditions of appropriate reliance. If a disclosure is not triggered because the update fell below the threshold, the provider bears documented responsibility for that classification, which can be reviewed in post-market monitoring.\footnote{While the stroke triage setting is chosen for its clarity and urgency, the same logic applies across other high-stakes AI deployments where accumulated user calibration is a safety-relevant asset: radiology reading aids, sepsis early-warning systems, credit risk models used by loan officers, or recidivism assessment tools used by parole boards. In each case, the problem of update opacity is not merely a technical question about model drift or performance maintenance; it is a governance question about how the epistemic conditions for appropriate reliance are preserved, documented, and communicated as the system evolves. Our framework provides a principled, operationalisable answer to this question.}

\subsection{Disclosure: What, How, and to Whom?}
\label{subsection:disclosure}
Our thresholding mechanism specifies \textit{when} an update becomes disclosure-relevant under continued admissibility, but it does not, by itself, settle \textit{what} information should be disclosed, nor \textit{how} it should be presented to different stakeholders. The relevant notion of disclosure is therefore indexed to a context of use: what should be disclosed, to whom, and in what form depends on the role-specific expectations, practical constraints, and governance responsibilities of the users involved. This is a familiar limitation of transparency work in xAI: the trigger rule can identify disclosure moments, yet the content and form of disclosure remain audience- and context-dependent \citep{hatherley2024virtues,hatherley2026defense,miller2019explanation,nyrup2022explanatory}. Addressing these questions requires moving from the formal machinery of threshold-based governance to the practical design of disclosure as a communicative act embedded in specific institutional and operational contexts.

For update opacity, the goal of disclosure is to communicate the aspects of change that may affect how outputs ought to be interpreted and relied upon following a maxim of selective relevance, not maximal disclosure. Here, information overload is itself a failure mode of transparency regimes \citep{hatherley2025clinicians}, particularly in settings where expert users operate under time and cognitive pressure and where the signal-to-noise ratio of technical disclosures is low \citep{zerilli2022transparency}. In practice, these aspects may include: (1) \textit{performance deltas} on agreed evaluation criteria, especially where subgroup-specific behaviour changes; (2) \textit{behavioural shift indicators}, such as changed recommendation frequencies, altered confidence distributions, or new instability at previously stable boundaries; (3) \textit{operating-envelope updates}, including changes in validated contexts of use or known failure modes; and (4) \textit{update rationale}, stated in non-technical terms, such as drift response, corrective maintenance, or infrastructure change. Disclosure must also be role-sensitive \citep{felzmann2019transparency}. Frontline users need concise, actionable information embedded in the workflow in which decisions are made. Governance leads and risk managers need structured summaries that support local oversight. Auditors and post-market reviewers need fuller documentation that links the update to validation evidence, rationale, and lifecycle records. These layers should be connected, but not collapsed into one artifact. A frontline user should not be burdened with technical annexes; an auditor should not be left with a generic banner and no traceable record.

Three general design principles follow. First, \textit{disclosure should be embedded in the operational context}: material change should be surfaced at the point of use when it is most likely to matter. Second, \textit{disclosure should be proportionate to role and urgency}: different users require different levels of detail and different channels. Third, \textit{disclosure should be acknowledgeable and traceable}: for accountability purposes, it should be possible to establish that relevant users or organizations received and acknowledged update-relevant information. This creates a closed epistemic loop: the provider discloses, the deployer receives and acknowledges, and the relevant users are informed in forms suited to their role. In this way, threshold-based disclosure becomes a practical device for preserving epistemic accessibility over time.

\section{Conclusions}
\label{section:conclusions}
Update opacity is a distinctively diachronic epistemic-accessibility problem arising from AI system change over time in a context of use. It arises when an AI system undergoes model-affecting system updates in ways that are no longer epistemically accessible enough for relevant users to understand how outputs should be interpreted and relied upon under their practical and institutional constraints. 
We argued that this problem cannot be solved either by indiscriminate disclosure or by purely internal operational monitoring. The AIA and MLOps each address part of the challenge. The AIA helps identify which kinds of change are normatively salient for continued trustworthy functioning; MLOps provides operational tools for tracking and comparing change over time. Our proposal combines these two governance cultures through trustworthiness profiles, trustworthiness levels, and threshold-based disclosure.
Our framework operationalises dynamic reporting and update compatibility without demanding the infeasible  disclosure of every update. More broadly, the framework suggests that governing AI system updates is not only a matter of technical maintenance or legal compliance, but of preserving the conditions under which changing systems remain intelligible enough to be appropriately used.
 Future work should refine domain-specific and formal choices of profile dimensions, dissimilarity functions, and thresholds, and test how different disclosure interfaces affect calibrated use in high-stakes settings.

\section*{Acknowledgments}
AF acknowledges partial support by the Swiss National Science Foundation (SNSF), grant no. 229061. JH thanks Marisha Mazumdar for her assistance constructing the medical AI example in section 5.3. JH also acknowledges support from Carlsgbergfondet, grant no. CF22-0243.


\bibliographystyle{plainnat} 
\bibliography{sample-base}

@techreport{EU_HLEG_TrustworthyAI_2019,
  author       = {{EU AI HLEG}},
  title        = {{Ethics Guidelines for Trustworthy AI}},
  year         = {2019},
  month        = apr,
  institution  = {European Commission},
  url          = {https://digital-strategy.ec.europa.eu/en/library/ethics-guidelines-trustworthy-ai}
}

@article{sparrow2020high,
  title={High hopes for “Deep Medicine”? {AI}, economics, and the future of care},
  author={Sparrow, Robert and Hatherley, Joshua},
  journal={Hastings Center Report},
  volume={50},
  number={1},
  pages={14--17},
  year={2020},
  publisher={Wiley Online Library}
}

@article{hatherley2024virtues,
  title={The virtues of interpretable medical {AI}},
  author={Hatherley, Joshua and Sparrow, Robert and Howard, Mark},
  journal={Cambridge Quarterly of Healthcare Ethics},
  volume={33},
  number={3},
  pages={323--332},
  year={2024},
  publisher={Cambridge University Press}
}

@article{sparrow2025continuous,
  title={Continuous learning as a threat to care},
  author={Sparrow, Robert and Daus, Zachary and Howard, Mark and Hatherley, Joshua and Kwan, Patrick},
  journal={Health and Technology},
  pages={1--5},
  year={2025},
  publisher={Springer}
}

@article{hatherley2026defense,
  title={In Defense of Post Hoc Explanations in Medical AI},
  author={Hatherley, Joshua and Aastrup Munch, Lauritz and Christian Bjerring, Jens},
  journal={Hastings Center Report},
  volume={56},
  number={1},
  pages={40--46},
  year={2026},
  publisher={Wiley Online Library}
}

@article{hatherley2025clinicians,
  title={Are clinicians ethically obligated to disclose their use of medical machine learning systems to patients?},
  author={Hatherley, Joshua},
  journal={Journal of medical ethics},
  volume={51},
  number={8},
  pages={567--573},
  year={2025},
  publisher={Institute of Medical Ethics}
}

@article{zerilli2022transparency,
  title={How transparency modulates trust in artificial intelligence},
  author={Zerilli, John and Bhatt, Umang and Weller, Adrian},
  journal={Patterns},
  volume={3},
  number={4},
  year={2022},
  publisher={Elsevier}
}

@article{felzmann2019transparency,
  title={Transparency you can trust: Transparency requirements for artificial intelligence between legal norms and contextual concerns},
  author={Felzmann, Heike and Villaronga, Eduard Fosch and Lutz, Christoph and Tam{\`o}-Larrieux, Aurelia},
  journal={Big Data \& Society},
  volume={6},
  number={1},
  pages={2053951719860542},
  year={2019},
  publisher={SAGE Publications Sage UK: London, England}
}

@article{miller2019explanation,
  title={Explanation in artificial intelligence: Insights from the social sciences},
  author={Miller, Tim},
  journal={Artificial intelligence},
  volume={267},
  pages={1--38},
  year={2019},
  publisher={Elsevier}
}

@article{nyrup2022explanatory,
  title={Explanatory pragmatism: a context-sensitive framework for explainable medical AI},
  author={Nyrup, Rune and Robinson, Diana},
  journal={Ethics and information technology},
  volume={24},
  number={1},
  pages={13},
  year={2022},
  publisher={Springer}
}

@article{ong2026considering,
  title={Considering the missing science of retraining and maintenance in medical artificial intelligence, using ophthalmology as an exemplar},
  author={Ong, Ariel Yuhan and Struyven, Robbert R and Denniston, Alastair K and Merle, David A and Engelmann, Justin and Kim, Hyunmin and Zhou, Yukun and Keane, Pearse A and Lains, Ines},
  journal={npj Digital Medicine},
  year={2026},
  publisher={Nature Publishing Group UK London}
}

@article{brown2021fda,
  title={FDA releases action plan for artificial intelligence/machine learning-enabled software as a medical device},
  author={Brown, Nathan A and Carey, Christin Helms and Gerry, Emily I},
  journal={The Journal of Robotics, Artificial Intelligence \& Law},
  volume={4},
  year={2021},
  publisher={HeinOnline}
}

@inproceedings{bansal2019updates,
  title={Updates in human-ai teams: Understanding and addressing the performance/compatibility tradeoff},
  author={Bansal, Gagan and Nushi, Besmira and Kamar, Ece and Weld, Daniel S and Lasecki, Walter S and Horvitz, Eric},
  booktitle={Proceedings of the AAAI conference on artificial intelligence},
  volume={33},
  number={01},
  pages={2429--2437},
  year={2019}
}

@techreport{EU_AI_Act_2024,
author = {{EU AI Act}},  
title        = {{Regulation (EU) 2024/1689 of the European Parliament and of the Council of 13 June 2024 laying down harmonised rules on artificial intelligence}},
  year         = {2024},
  month        = jun,
  institution  = {European Union},
  howpublished = {\emph{Official Journal of the European Union}, L 1689, 12 July 2024},
   url          = {https://eur-lex.europa.eu/eli/reg/2024/1689/oj}
}

@book{Bradford2020BrusselsEffect,
  author    = {Bradford, Anu},
  title     = {The {B}russels Effect},
  publisher = {Oxford University Press},
  year      = {2020},
  address   = {Oxford, UK}
}

@article{ferrario2024justifying,
  title={Justifying our credences in the trustworthiness of {AI} systems: {A} reliabilistic approach},
  author={Ferrario, Andrea},
  journal={Science and Engineering Ethics},
  volume={30},
  number={6},
  pages={55},
  year={2024},
  publisher={Springer}
}

@inproceedings{ferrario2025trustworthiness,
  title={A Trustworthiness-based metaphysics of {a}rtificial {i}ntelligence systems},
  author={Ferrario, Andrea},
  booktitle={{Proceedings of the 2025 ACM Conference on Fairness, Accountability, and Transparency}},
  pages={1360--1370},
  year={2025}
}

@article{hatherley2025moving,
  title={A moving target in {AI}-assisted decision-making: {D}ataset shift, model updating, and the problem of update opacity},
  author={Hatherley, Joshua},
  journal={{Ethics and Information Technology}}
            ,
  volume={27},
  number={2},
  pages={1--14},
  year={2025},
  publisher={Springer}
}

@article{hatherley2023diachronic,
  title={Diachronic and synchronic variation in the performance of adaptive machine learning systems: {T}he ethical challenges},
  author={Hatherley, Joshua and Sparrow, Robert},
  journal={{Journal of the American Medical Informatics Association}},
  volume={30},
  number={2},
  pages={361--366},
  year={2023},
  publisher={Oxford University Press}
}

@inproceedings{facchini2021towards,
  title={Towards a taxonomy for the opacity of {AI} systems},
  author={Facchini, Alessandro and Termine, Alberto},
  booktitle={{Conference on Philosophy and Theory of Artificial Intelligence}},
  pages={73--89},
  year={2021},
  organization={Springer}
}

@article{gilbert2021algorithm,
  title={Algorithm change protocols in the regulation of adaptive machine learning--based medical devices},
  author={Gilbert, Stephen and Fenech, Matthew and Hirsch, Martin and Upadhyay, Shubhanan and Biasiucci, Andrea and Starlinger, Johannes},
  journal={Journal of Medical Internet Research},
  volume={23},
  number={10},
  pages={e30545},
  year={2021},
  publisher={JMIR Publications Toronto, Canada}
}

@article{kreuzberger2023machine,
  title={Machine learning operations ({MLOps}): {O}verview, definition, and architecture},
  author={Kreuzberger, Dominik and K{\"u}hl, Niklas and Hirschl, Sebastian},
  journal={IEEE Access},
  volume={11},
  pages={31866--31879},
  year={2023},
  publisher={IEEE}
}

@article{bayram2024adaptive,
  title={Adaptive data quality scoring operations framework using drift-aware mechanism for industrial applications},
  author={Bayram, Firas and Ahmed, Bestoun S and Hallin, Erik},
  journal={Journal of Systems and Software},
  volume={217},
  pages={112184},
  year={2024},
  publisher={Elsevier}
}

@article{liu2013change,
  title={Change-point detection in time-series data by relative density-ratio estimation},
  author={Liu, Song and Yamada, Makoto and Collier, Nigel and Sugiyama, Masashi},
  journal={Neural Networks},
  volume={43},
  pages={72--83},
  year={2013},
  publisher={Elsevier}
}

@article{liu2022concept,
  title={Concept drift detection delay index},
  author={Liu, Anjin and Lu, Jie and Song, Yiliao and Xuan, Junyu and Zhang, Guangquan},
  journal={IEEE Transactions on Knowledge and Data Engineering},
  volume={35},
  number={5},
  pages={4585--4597},
  year={2022},
  publisher={IEEE}
}

@article{lu2018learning,
  title={Learning under concept drift: {A} review},
  author={Lu, Jie and Liu, Anjin and Dong, Fan and Gu, Feng and Gama, Joao and Zhang, Guangquan},
  journal={IEEE Transactions on Knowledge and Data Engineering},
  volume={31},
  number={12},
  pages={2346--2363},
  year={2018},
  publisher={IEEE}
}

@article{ferrario2022robustness,
  title={The robustness of counterfactual explanations over time},
  author={Ferrario, Andrea and Loi, Michele},
  journal={IEEE Access},
  volume={10},
  pages={82736--82750},
  year={2022},
  publisher={IEEE}
}

@article{guidotti2018survey,
  title={A survey of methods for explaining black box models},
  author={Guidotti, Riccardo and Monreale, Anna and Ruggieri, Salvatore and Turini, Franco and Giannotti, Fosca and Pedreschi, Dino},
  journal={ACM Computing Surveys (CSUR)},
  volume={51},
  number={5},
  pages={1--42},
  year={2018},
  publisher={ACM New York, NY, USA}
}

@article{eken2025multivocal,
  title={A multivocal review of {MLOps} practices, challenges and open issues},
  author={Eken, Beyza and Pallewatta, Samodha and Tran, Nguyen and Tosun, Ayse and Babar, Muhammad Ali},
  journal={ACM Computing Surveys},
  volume={58},
  number={2},
  pages={1--35},
  year={2025},
  publisher={ACM New York, NY}
}

@inproceedings{barocas2020hidden,
  title={The hidden assumptions behind counterfactual explanations and principal reasons},
  author={Barocas, Solon and Selbst, Andrew D and Raghavan, Manish},
  booktitle={Proceedings of the 2020 Conference on Fairness, Accountability, and Transparency},
  pages={80--89},
  year={2020}
}

@inproceedings{fonseca2023setting,
  title={Setting the right expectations: Algorithmic recourse over time},
  author={Fonseca, Jo{\~a}o and Bell, Andrew and Abrate, Carlo and Bonchi, Francesco and Stoyanovich, Julia},
  booktitle={Proceedings of the 3rd ACM Conference on Equity and Access in Algorithms, Mechanisms, and Optimization},
  pages={1--11},
  year={2023}
}

@inproceedings{venkatasubramanian2020philosophical,
  title={The philosophical basis of algorithmic recourse},
  author={Venkatasubramanian, Suresh and Alfano, Mark},
  booktitle={Proceedings of the 2020 Conference on Fairness, Accountability, and Transparency},
  pages={284--293},
  year={2020}
}

@article{sogaard2023opacity,
  title={On the opacity of deep neural networks},
  author={S{\o}gaard, Anders},
  journal={Canadian Journal of Philosophy},
  volume={53},
  number={3},
  pages={224--239},
  year={2023},
  publisher={Cambridge University Press}
}

@article{rabanser2026towards,
  title={Towards a science of {AI} agent reliability},
  author={Rabanser, Stephan and Kapoor, Sayash and Kirgis, Peter and Liu, Kangheng and Utpala, Saiteja and Narayanan, Arvind},
  journal={arXiv preprint arXiv:2602.16666},
  year={2026}
}

@techreport{NIST_AIRMF_2023,
  author      = {{National Institute of Standards and Technology}},
  title       = {Artificial Intelligence Risk Management Framework (AI RMF 1.0)},
  institution = {U.S. Department of Commerce, National Institute of Standards and Technology},
  year        = {2023},
  number      = {NIST AI 100-1},
  doi         = {10.6028/NIST.AI.100-1},
  url         = {https://nvlpubs.nist.gov/nistpubs/ai/nist.ai.100-1.pdf}
}

@techreport{OECD_TrustworthyAI_2021,
  author      = {{OECD}},
  title       = {Tools for Trustworthy AI},
  institution = {Organisation for Economic Co-operation and Development},
  year        = {2021},
  url         = {https://www.oecd.org/content/dam/oecd/en/publications/reports/2021/06/tools-for-trustworthy-ai_0e36bb08/008232ec-en.pdf}
}

@misc{ferrario2026highriskidentity,
  title         = {{High-Risk AI systems and the problem of identity in the European AI Act}},
  author        = {Ferrario, Andrea},
  year          = {2026},
  note          = {Accepted as a non-archival paper at \emph{The 2026 ACM Conference on Fairness, Accountability, and Transparency (FAccT '26)}, Montreal, QC, Canada, June 25--28, 2026}
}

\end{document}